\documentclass[conference,twocolumn,10pt]{IEEEtran}

\pdfoutput=1
\usepackage[cmex10]{amsmath}
\usepackage{amsfonts,amssymb}
\usepackage{verbatim}
\usepackage{color}
\usepackage{multirow} 

\usepackage{cite}

\ifCLASSINFOpdf
  \usepackage[pdftex]{graphicx}
  \graphicspath{{graphics/}}
  \DeclareGraphicsExtensions{.pdf,.jpeg,.png}
\else
  \usepackage[dvips]{graphicx}
  \graphicspath{{graphics/}}
  \DeclareGraphicsExtensions{.eps}
\fi

\usepackage{array}

\usepackage{mdwmath}
\usepackage{mdwtab}
\DeclareMathOperator{\erf}{erf}

\begin{document}
\bibliographystyle{IEEEtran}
%
\title{Active Versus Passive: Receiver Model Transforms for Diffusive Molecular Communication}

\author{\IEEEauthorblockN{Adam Noel$^{\ast\dagger}$, Yansha Deng$^{\ddagger}$, Dimitrios Makrakis$^{\dagger}$, and Abdelhakim Hafid$^{\ast}$}
\IEEEauthorblockA{$^{\ast}$Department of Computer Science and Operations Research,
	University of  Montreal
	\\ $^{\dagger}$School of Electrical Engineering and Computer Science,
	University of Ottawa
\\ $^{\ddagger}$Department of Informatics,
King's College London}}


\newcommand{\specialcell}[2][l]{\begin{tabular}[#1]{@{}l@{}}#2\end{tabular}}

\newcommand{\x}{x}
\newcommand{\y}{y}
\newcommand{\z}{z}

\newcommand{\second}{\textnormal{s}}
\newcommand{\metre}{\textnormal{m}}

\newcommand{\EDAvgt}[1]{\overline{\textnormal{ED}}\left(#1\right)}
\newcommand{\EDAvg}{\overline{\textnormal{ED}}}
\newcommand{\EDt}[1]{\textnormal{ED}\left(#1\right)}
\newcommand{\Scaling}{\mathcal{S}}
\newcommand{\threeD}{\textnormal{3D}}
\newcommand{\oneD}{\textnormal{1D}}
\newcommand{\pass}{\textnormal{PA}}
\newcommand{\absorb}{\textnormal{AB}}
\newcommand{\mol}{\textnormal{mol}}

\newcommand{\kth}[1]{k_{#1}}

\newcommand{\M}{M}
\newcommand{\T}{T}
\newcommand{\At}{A(\tx{})}
\newcommand{\Dx}[1]{D_{#1}}
\newcommand{\Nx}[1]{N_{#1}}
\newcommand{\Nxtavg}[2]{\overline{\Nx{}}_\textnormal{#2}\left(#1\right)}
\newcommand{\NxAvg}[1]{\overline{\Nx{}}_\textnormal{#1}}
\newcommand{\Cxtavg}[2]{\overline{C}_\textnormal{#2}\left(#1\right)}
\newcommand{\Nxt}[2]{{\Nx{}}_\textnormal{#2}\left(#1\right)}

\newcommand{\EXP}[1]{\exp\left(#1\right)}
\newcommand{\ERF}[1]{\erf\left(#1\right)}
\newcommand{\ERFC}[1]{\mathrm{erfc}\left(#1\right)}


\newcommand{\Vol}[1]{V_{#1}}
\newcommand{\subV}[1]{S_{#1}}
\newcommand{\dt}[1]{\Delta\tx{\textnormal{#1}}}
\newcommand{\prop}[1]{a_{#1}}
\newcommand{\molNum}[1]{U_{#1}}
\newcommand{\rx}{\textnormal{RX}}
\newcommand{\rrx}{r_\textnormal{RX}}
\newcommand{\Vrx}{V_\textnormal{RX}}
\newcommand{\dist}{d}
\newcommand{\tx}[1]{t_{#1}}
\newcommand{\kx}[1]{k_{#1}}
\newcommand{\wx}[1]{w_{#1}}
\newcommand{\thresh}{\xi}

\newcommand{\weight}[1]{w_{#1}}

\newcommand{\new}[1]{\textbf{#1}}

\newcommand{\smM}{m}

\newcommand{\edit}[2]{#1}

\maketitle

\begin{abstract}
This paper presents an analytical comparison of active and passive receiver models in diffusive molecular communication. In the active model, molecules are absorbed when they collide with the receiver surface. In the passive model, the receiver is a virtual boundary that does not affect molecule behavior. Two approaches are presented to derive transforms between the receiver signals. As an example, two models for an unbounded diffusion-only molecular communication system with a spherical receiver are unified. As time increases in the three-dimensional system, the transform functions have constant scaling factors, such that the receiver models are effectively equivalent. Methods are presented to enable the transformation of stochastic simulations, which are used to verify the transforms and demonstrate that transforming the simulation of a passive receiver can be more efficient and more accurate than the direct simulation of an absorbing receiver.
\end{abstract}

\section{Introduction}

Molecular communication has been receiving increasing attention as a strategy for the design of novel communication systems in fluid environments; see \cite{Farsad2016}. Much of this attention has been focused on molecular communication via diffusion, whose attractive properties include no required external energy or infrastructure for the molecules to propagate from a transmitter. Once molecules are released by the transmitter, they move in the fluid via a process that is effectively random.

Arguably one of the biggest schisms in the analysis of communication via diffusion is in the modeling of the receiver. Receiver models are generally classified as either \emph{passive} or \emph{active}. A passive receiver can observe but has no effect on molecule behavior. An active receiver is typically a site for chemical reactions, either inside or on its surface, and molecules can be identified when a reaction takes place. Passive receiver models are commonly favored for their simplicity in analysis and simulation, whereas active models are more realistic in representing the chemical detection of molecules.

The vast majority of studies of single communication links assumes a receiver model and then proceeds without assessing the choice. For example, active receivers have been considered in \cite{Chou2015b,Movahednasab2016}, and passive receivers have been considered in \cite{Meng2014} and the first author's work in \cite{Noel2014d}. Explicit comparisons between receiver models are very uncommon; one example is \cite{Farsad2013}, where the one-dimensional passive and absorbing models were both fitted to experimental data obtained from the tabletop testbed in \cite{Farsad2013a}. The authors observed similarities between the two models and attributed them to the dominance of airflow. Other authors, such as in \cite{Mosayebi2014,Jamali2015}, derived results that were valid for any model but did not compare models directly.

In this work, we compare active and passive receiver models. Our goal is to unify the models so that we can \emph{transform} signals from one model to the other. As an example, we unify the channel impulse responses (CIRs) for a spherical receiver in an unbounded diffusion-only environment. Other realistic phenomena, such as fluid flow or the potential for other chemical reactions, may be considered in future work. We consider both a \emph{perfectly-absorbing} receiver and a \emph{passive} receiver in three dimensions (3D) and in one dimension (1D, i.e., the receiver is treated as a line segment). All of these systems have been studied extensively, and we omit the two-dimensional environment because the CIR of the absorbing circle has not been described in closed form; see \cite{Farsad2016}. By carefully choosing how to represent the CIRs, we derive functions that transform a signal at a passive receiver into a signal at the corresponding absorbing receiver. In particular, we make the following contributions:
\begin{enumerate}
	\item We take the integral of the passive CIR so that it includes prior signal information and is then comparable to the active CIR. In practice, this can be measured using the weighted sum detector that we introduced in \cite{Noel2014d}. Alternatively, we take the derivative of the active CIR so that it is an instantaneous measurement to compare with the passive CIR. The derivative can be obtained from the hitting rate detector proposed in \cite{Heren2015}. These two approaches, which lead us to derive transform functions for the 1D and 3D environments, apply to absorbing and passive receivers in \emph{any} environment.
	\item We describe the proper implementation of the energy detector at a passive receiver and the hitting rate detector at an absorbing receiver to perform the integral and derivative operations, respectively. These detector modifications are needed to accurately apply the transforms between active and passive signals in simulations.
	\item We demonstrate the accuracy of the transforms and their inverses, both analytically and via simulation. As a result, we can work with whichever model is currently most appropriate and then accurately transform the signal to the other model.
\end{enumerate}

By deriving transforms between the passive and the perfectly-absorbing receiver models for the sphere in a diffusion-only system, we effectively unify \emph{all} existing literature that has selected one of these models. This unification makes the initial selection of a receiver model less critical. We can analyze one model and then simulate the other model, or divide the analysis between the two models, as desired, which gives us greater flexibility when deciding which model to use. As an example, we show that it can be both more accurate \emph{and} more efficient to simulate an absorbing receiver signal by transforming a passive receiver simulation instead of directly simulating the absorbing receiver. This is due to the latter's requirement of a very small simulation time step for accuracy.

We also note that our analysis focuses on \emph{expected} CIRs. The impact of the transform functions on the channel statistics will be considered in future work. However, we verify the transforms numerically and also using average results obtained from our particle-based stochastic molecular communications simulator AcCoRD (Actor-based Communication via Reaction-Diffusion), which is available as a public beta \cite{Noel2016}.

The rest of this paper is organized as follows. Section~\ref{sec_model} describes the system models and presents the corresponding CIRs. In Section~\ref{sec_unifying}, we derive the transform functions. In Section~\ref{sec_implementation}, we discuss the implementation of the integral and derivative operations. We verify our analytical results with simulations in Section~\ref{sec_results}, and conclude in Section~\ref{sec_concl}.

\section{System Model and Channel Impulse Responses}
\label{sec_model}

We consider a point transmitter (TX) releasing $\Nx{}$ molecules into an unbounded 1D or 3D environment. We assume uniform temperature and viscosity, and that the local molecule concentration is sufficiently low, so that molecules diffuse with constant diffusion coefficient $\Dx{}$. These molecules are observed by a receiver (RX) that is centered at a distance $\dist$ from the TX and has radius $\rrx$, i.e., a segment of length $2\rrx$ in 1D and a sphere in 3D. If the RX is active, then we consider a perfectly-absorbing surface that removes and counts molecules as they arrive, i.e., the absorption rate is $\kx{} \to \infty$. If the RX is passive, then it has no impact on molecule behavior but is able to count the number of molecules within its virtual boundary at any instant. Finally, we assume that diffusion is the only phenomenon affecting molecule behavior in the propagation environment (except for absorption at the active RX's surface).

We define the channel impulse response (CIR) $\Nxtavg{t}{RX}$ as the number of molecules \emph{expected} at the RX at time $\tx{}$, given that $\Nx{}$ molecules are instantaneously released by the TX at time $\tx{}=0$. In the remainder of this section, we present the CIRs that are used throughout the remainder of this paper.

\subsection{3D Channel Impulse Responses}

If the environment is 3D, then the CIR of the absorbing RX is given by \cite[Eq.~(23)]{Yilmaz2014b}
\begin{equation}
\Nxtavg{t}{RX}|^\absorb_\threeD = \frac{\Nx{}\rrx}{\dist}\ERFC{\frac{\dist-\rrx}{\sqrt{4\Dx{}\tx{}}}},
\label{absorbing_response}
\end{equation}
where $\ERFC{x} = 1 - \ERF{x}$ is the complementary error function (from \cite[Eq.~(8.250.4)]{Gradshteyn2007}), and the ``$\absorb$'' superscript means ``absorbing''. Eq.~(\ref{absorbing_response}) describes the \emph{total} number of molecules absorbed by time $\tx{}$. For the passive RX, the expected \emph{point} concentration $\Cxtavg{t}{point}$ is \cite[Eq.~(4.28)]{Crank1979}
\begin{equation}
\Cxtavg{t}{point}|^\pass_\threeD = \frac{\Nx{}}{(4\pi \Dx{}
	\tx{})^{3/2}}\EXP{-\frac{\dist^2}{4\Dx{}\tx{}}},
\label{passive_response_point}
\end{equation}
where we use the ``$\pass$'' superscript to denote ``passive''.


It is common to assume that the molecule concentration is uniform inside a passive RX, which is justified if the RX is sufficiently far from the TX, i.e., if $\dist \gg \rrx$ (as we demonstrated in \cite{Noel2013b}). Here, we make this assumption for ease of analysis. Thus, we multiply (\ref{passive_response_point}) by the RX volume $\Vrx$ to write the number of molecules expected inside the RX as
\begin{equation}
\Nxtavg{t}{RX}|^\pass_\threeD = \frac{\Nx{}\Vrx}{(4\pi \Dx{}
	\tx{})^{3/2}}\EXP{-\frac{\dist^2}{4\Dx{}\tx{}}}.
\label{passive_response_UCA}
\end{equation}

\subsection{1D Channel Impulse Responses}

If the environment is 1D, then the CIR of the absorbing RX is given by \cite[Eq.~(7)]{Farsad2016}
\begin{equation}
\Nxtavg{t}{RX}|^\absorb_\oneD = \Nx{}\ERFC{\frac{\dist-\rrx}{\sqrt{4\Dx{}\tx{}}}}.
\label{absorbing_response_1D}
\end{equation}

The expected \emph{point} concentration for the 1D passive RX is \cite[Eq.~(3.6)]{Crank1979}
\begin{equation}
\Cxtavg{t}{point}|^\pass_\oneD = \frac{\Nx{}}{\sqrt{4\pi \Dx{}
	\tx{}}}\EXP{-\frac{\dist^2}{4\Dx{}\tx{}}}.
\label{passive_response_point_1D}
\end{equation}


If we assume that the passive RX is sufficiently far from the TX, then the simplified 1D CIR is directly from (\ref{passive_response_point_1D}) as
\begin{equation}
\Nxtavg{t}{RX}|^\pass_\oneD = \frac{\rrx\Nx{}}{\sqrt{\pi \Dx{}
	\tx{}}}\EXP{-\frac{\dist^2}{4\Dx{}\tx{}}}.
\label{passive_response_UCA_1D}
\end{equation}


\section{Unifying Receiver Models}
\label{sec_unifying}

In this section, we seek the existence of transform functions that have the form
\begin{equation}
\textnormal{Absorbing Signal} \stackrel{?}{=} \Scaling(\textnormal{Passive Signal}),
\label{transform}
\end{equation}
where we emphasize that we are most interested in obtaining an active RX signal from a passive RX, since a passive RX is generally faster to simulate (though simulating absorption can be faster \emph{if} most molecules get absorbed). We do not constrain the ``signals'' in (\ref{transform}) to be CIRs; rather, we seek to manipulate either the absorbing or the passive CIR so that it is \emph{comparable} to the other.

We claim that the CIRs of the two receiver models can be perceived as similar measurements but they provide fundamentally different information about what has happened at the RX. A sample of the absorbing RX's CIR is the \emph{total} number of molecules that have arrived at that RX since they were released by the TX, i.e., a sample includes history information. However, a sample of the passive RX's CIR is \emph{only} the current number of molecules that are inside the RX at the instant when the sample is taken; the history of the molecules that have entered and left the passive RX is ignored. So, to derive a transform, we propose using either a passive receiver model that accounts for signal ``history'', or an active receiver model that describes the instantaneous behavior.

We first consider the 3D environment, where we derive the transform function that applies to the passive signal with history information, $\Scaling'_\threeD$, and that which applies to the instantaneous passive signal, $\Scaling''_\threeD$. These transforms will simplify to \emph{constant} scaling factors in the asymptotic case, i.e., as $\tx{}\to\infty$, which may be useful for modeling intersymbol interference (ISI). Then, we consider the 1D environment, where the corresponding transform functions are $\Scaling'_\oneD$ and $\Scaling''_\oneD$.

\subsection{3D Analysis}

We begin our 3D analysis with the more intuitive signal manipulation, which is to model the receiver as an energy detector. The CIR of the absorbing RX is intuitively a measure of the received energy over time, since the signal accounts for every molecule arrival. Analytically, an energy detector for the passive RX is defined by integrating the CIR over time, and we will show in Section~\ref{sec_implementation} that this can be implemented with a weighted sum detector. Analogously to \cite[Eq.~(3.5b)]{Crank1979}, which derives the instantaneous signal due to a point source that is continuously releasing molecules, we can integrate (\ref{passive_response_UCA}) over $\tx{}$ to write the energy detector signal $\EDAvgt{\tx{}}$ as
\begin{equation}
\EDAvgt{\tx{}}|^\pass_\threeD = \frac{\Nx{}\Vrx}{4\pi\Dx{}\dist}\ERFC{\frac{\dist}{\sqrt{4\Dx{}\tx{}}}}.
\label{passive_response_integral}
\end{equation}

We can immediately compare (\ref{passive_response_integral}) with the absorbing CIR in (\ref{absorbing_response}), even though there is an abuse of notation since (\ref{absorbing_response}) has unit $[\mol]$ (i.e., molecule) whereas (\ref{passive_response_integral}) has units $[\mol\cdot\second]$. This difference is a side effect of having physically different RXs. To write (\ref{passive_response_integral}) as a function of (\ref{absorbing_response}), we need a way to separate the terms inside the complementary error function in (\ref{absorbing_response}). To do this, we will use the elementary approximation of $\ERF{x}$ in \cite[Eq.~(4a)]{Lether1993}, which we have observed to have a relative error of less than $1\%$ for $0 \le x \le 2.5$, to re-write $\ERFC{x}$ as
\begin{equation}
\ERFC{x} \approx \EXP{-\frac{16}{23}x^2 - \frac{2}{\sqrt{\pi}}x},
\end{equation}
and therefore write $\ERFC{\cdot}$ in (\ref{absorbing_response}) as
\begin{equation}
\ERFC{\frac{\dist-\rrx}{\sqrt{4\Dx{}\tx{}}}} \approx \ERFC{\frac{\dist}{\sqrt{4\Dx{}\tx{}}}}\At,
\label{erfc_approx}
\end{equation}
where we define the function $\At$ as
\begin{equation}
\At = \EXP{\frac{\rrx}{\sqrt{\Dx{}\tx{}}}\left(\frac{4(2\dist-\rrx)}{23\sqrt{\Dx{}\tx{}}}+\frac{1}{\sqrt{\pi}}\right)},
\label{erfc_approx_helper}
\end{equation}
and in practice (\ref{erfc_approx}) is very accurate unless $\tx{}$ is very small.
Using (\ref{erfc_approx}) and the equation for the volume of a sphere, we write the transform function $\Scaling'_\threeD$ as
\begin{equation}
\Nxtavg{\tx{}}{RX}|^\absorb_\threeD = \Scaling'_\threeD(\EDAvgt{\tx{}}|^\pass_\threeD)
\approx \frac{3\Dx{}\At}{\rrx^2} \EDAvgt{\tx{}}|^\pass_\threeD,
\label{transform_3D}
\end{equation}
and we can re-arrange (\ref{transform_3D}) to find the inverse transform $\Scaling'^{-1}_\threeD$.

In the asymptotic case, i.e., as $\tx{} \to \infty$, the complementary error function goes to 1 and thus the transform function $\Scaling'^\infty_\threeD$ has a \emph{constant} scaling factor
\begin{equation}
\Nxtavg{\tx{}}{RX}|^\absorb_\threeD = \Scaling'^\infty_\threeD(\EDAvgt{\tx{}}|^\pass_\threeD)
\approx \frac{3\Dx{}}{\rrx^2} \EDAvgt{\tx{}}|^\pass_\threeD.
\label{transform_3D_asym}
\end{equation}

From (\ref{transform_3D_asym}), we see that these two fundamentally different receiver models are effectively \emph{equivalent} (asymptotically).

Next, we perform the complementary signal manipulation, i.e., we seek a measure of the instantaneous behavior of the absorbing receiver model to compare with the passive RX CIR. Thus, we take the derivative of the active RX's CIR. This is the rate of molecule absorption at the RX, $\Delta\Nxtavg{t}{RX}$, and has been previously presented as \cite[Eq.~(22)]{Yilmaz2014b}
\begin{equation}
\Delta\Nxtavg{t}{RX}|^\absorb_\threeD = \frac{\Nx{}\rrx(\dist-\rrx)}{\dist\sqrt{4\pi \Dx{}\tx{}^3}		}\EXP{-\frac{(\dist-\rrx)^2}{4\Dx{}\tx{}}}.
\label{absorbing_response_derivative}
\end{equation}

The implementation of (\ref{absorbing_response_derivative}) in simulations will be discussed in Section~\ref{sec_implementation}. Here, we can compare (\ref{absorbing_response_derivative}) with (\ref{passive_response_UCA}) (once again with a slight abuse of notation since (\ref{absorbing_response_derivative}) is in $[\mol\cdot\second^{-1}]$ and (\ref{passive_response_UCA}) is in $[\mol]$). From the properties of exponential functions, it can be shown that the transform function $\Scaling''_\threeD$ is
\begin{align}
\Delta\Nxtavg{t}{RX}|^\absorb_\threeD = &\,
\Scaling''_\threeD(\Nxtavg{t}{RX}|^\pass_\threeD) \nonumber \\
= &\, \frac{3\Dx{}(\dist-\rrx)}{\rrx^2\dist}\Nxtavg{t}{RX}|^\pass_\threeD \nonumber \\
& \times \EXP{\frac{\rrx(2\dist-\rrx)}{4\Dx{}\tx{}}},
\label{transform_3D_instant}
\end{align}

Asymptotically, the transform function $\Scaling''^\infty_\threeD$ has a \emph{constant} scaling factor, i.e.,
\begin{align}
\Delta\Nxtavg{t}{RX}|^\absorb_\threeD = &\,
\Scaling''^\infty_\threeD(\Nxtavg{t}{RX}|^\pass_\threeD) \nonumber \\
\approx &\,\frac{3\Dx{}(\dist-\rrx)}{\rrx^2\dist}\Nxtavg{t}{RX}|^\pass_\threeD,
\label{transform_3D_instant_asym}
\end{align}
and if we again assume that the TX is sufficiently far from the RX (i.e., $\dist \gg \rrx$), then the two asymptotic transform functions $\Scaling'^\infty_\threeD$ and $\Scaling''^\infty_\threeD$ are analogous for the 3D receiver models with the \emph{same} scaling factor, i.e., $3\Dx{}/\rrx^2$.

\subsection{1D Analysis}

Our strategy to transform between the active and passive receiver models in the 1D system is the same as that applied for the 3D system. The energy detector signal for the 1D system can be obtained by integrating the passive CIR in (\ref{passive_response_UCA_1D}) over time. If we use $\tau$ as the dummy variable of integration over time, then we can integrate (\ref{passive_response_UCA_1D}) via the substitution $x = \tau^{-\frac{1}{2}}$, the integral \cite[Eq.~(2.325.12)]{Gradshteyn2007}
\begin{align}
\int \frac{1}{x^2}\EXP{-ax^2}\mathrm{d}x = -\frac{\EXP{-ax^2}}{x} + \sqrt{a\pi}\ERF{-\sqrt{a}x},
\end{align}
and the definition of $\ERFC{x}$. By performing these steps, we derive the expected energy detector signal at the 1D passive RX as
\begin{align}
\EDAvgt{\tx{}}|^\pass_\oneD = &\, 2\rrx\Nx{}\bigg[\sqrt{\frac{\tx{}}{\pi\Dx{}}}\EXP{-\frac{\dist^2}{4\Dx{}\tx{}}} \nonumber \\
& -\frac{\dist}{2\Dx{}}\ERFC{\frac{\dist}{\sqrt{4\Dx{}\tx{}}}}\bigg].
\label{passive_response_integral_1D}
\end{align}

We compare (\ref{passive_response_integral_1D}) with the 1D absorbing CIR in (\ref{absorbing_response_1D}). We do so to obtain the transform function $\Scaling'_\oneD$ as
\begin{align}
\Nxtavg{\tx{}}{RX}|^\absorb_\oneD = &\, \Scaling'_\oneD(\EDAvgt{\tx{}}|^\pass_\oneD) \nonumber \\
= &\,\frac{2\Dx{}\At}{\dist}\bigg[\Nx{}\sqrt{\frac{\tx{}}{\pi\Dx{}}}\EXP{-\frac{\dist^2}{4\Dx{}\tx{}}} \nonumber \\
&\, - \frac{\EDAvgt{\tx{}}|^\pass_\oneD}{2\rrx}\bigg],
\label{transform_1D}
\end{align}
where $\At$ is in (\ref{erfc_approx_helper}). As $\tx{} \to \infty$, the transform becomes
\begin{align}
\Nxtavg{\tx{}}{RX}|^\absorb_\oneD = &\, \Scaling'^\infty_\oneD(\EDAvgt{\tx{}}|^\pass_\oneD) \nonumber \\
\approx &\,\frac{2\Dx{}}{\dist}\bigg[\Nx{}\sqrt{\frac{\tx{}}{\pi\Dx{}}} - \frac{\EDAvgt{\tx{}}|^\pass_\oneD}{2\rrx}\bigg],
\label{transform_1D_asym}
\end{align}
which does \emph{not} have a constant scaling factor.

Finally, we determine the transform function $\Scaling''_\oneD$. The derivative of the CIR at the 1D absorbing RX in (\ref{absorbing_response_1D}) is the rate of molecule absorption at, and from \cite[Eq.~(6)]{Farsad2016} is
\begin{equation}
\Delta\Nxtavg{t}{RX}|^\absorb_\oneD = \frac{\Nx{}(\dist-\rrx)}{\sqrt{4\pi \Dx{}\tx{}^3}}
\EXP{-\frac{(\dist-\rrx)^2}{4\Dx{}\tx{}}}.
\label{absorbing_response_derivative_1D}
\end{equation}

We can compare (\ref{absorbing_response_derivative_1D}) with (\ref{passive_response_UCA_1D}) to derive the transform function $\Scaling''_\oneD$ as
\begin{align}
\Delta\Nxtavg{t}{RX}|^\absorb_\oneD = &\, \Scaling''_\oneD(\Nxtavg{t}{RX}|^\pass_\oneD) \nonumber \\
= &\, \frac{\dist-\rrx}{2\rrx\tx{}}\Nxtavg{t}{RX}|^\pass_\oneD
\EXP{\frac{\rrx(2\dist-\rrx)}{4\Dx{}\tx{}}},
\label{transform_1D_instant}
\end{align}
which closely resembles its 3D variant $\Scaling''_\threeD$ in (\ref{transform_3D_instant}). However, one key difference is that it does \emph{not} have a constant scaling factor in the asymptotic case, i.e.,
\begin{equation}
\Delta\Nxtavg{t}{RX}|^\absorb_\oneD = \Scaling''^\infty_\oneD(\Nxtavg{t}{RX}|^\pass_\oneD)
\approx \frac{\dist-\rrx}{2\rrx\tx{}}\Nxtavg{t}{RX}|^\pass_\oneD.
\label{transform_1D_instant_asym}
\end{equation}

The 1D and 3D environments that we considered have very similar configurations but resulted in different transform functions to convert between the active and passive receiver models. For reference, we have summarized the equations for the signals and their transforms in Table~\ref{table_equations}. We note that the effective equivalence (via the constant scaling factor $3\Dx{}/\rrx^2$) that we observed for the 3D model in the asymptotic case was not observed for the 1D model. Even though each transform function only applies to its corresponding system model, our approach to derive the transform functions could be applied to \emph{any} system model where comparable CIRs can be found.

\begin{table}[!tb]
	\centering
	\caption{Summary of analytical expressions, which are sorted according to the labeled curves in the figures in Section~\ref{sec_results}. Equations denoted with ``$\ast$'' are not shown in any figure.}
	
	{\renewcommand{\arraystretch}{1.3}
		\begin{tabular}{l|c||c|c}
			\hline
			\bfseries Curve & \bfseries Signal & \bfseries System 1 & \bfseries System 2 \\ \hline \hline
			\multirow{4}{*}{Analytical} & Passive CIR
			& (\ref{passive_response_UCA}) & (\ref{passive_response_UCA_1D}) \\ \cline{2-4}
			& Passive ED
			& (\ref{passive_response_integral}) & (\ref{passive_response_integral_1D})$^\ast$ \\ \cline{2-4}
			& Absorbing CIR
			& (\ref{absorbing_response}) & (\ref{absorbing_response_1D})$^\ast$ \\ \cline{2-4}
			& Absorption Rate
			& (\ref{absorbing_response_derivative}) & (\ref{absorbing_response_derivative_1D}) \\ \hline
			\multirow{4}{*}{\specialcell{Analytical\\via Transform}} & Passive CIR
			& (\ref{transform_3D_instant})$^{-1}$ & (\ref{transform_1D_instant})$^{-1}$ \\ \cline{2-4}
			& Passive ED
			& (\ref{transform_3D})$^{-1}$ & (\ref{transform_1D})$^{-1\ast}$ \\ \cline{2-4}
			& Absorbing CIR
			& (\ref{transform_3D}) & (\ref{transform_1D})$^\ast$ \\ \cline{2-4}
			& Absorption Rate
			& (\ref{transform_3D_instant}) & (\ref{transform_1D_instant}) \\ \hline
			\multirow{4}{*}{\specialcell{Asymptotic\\Transform Analytical}} & Passive CIR
			& (\ref{transform_3D_instant_asym})$^{-1}$ & (\ref{transform_1D_instant_asym})$^{-1}$ \\ \cline{2-4}
			& Passive ED
			& (\ref{transform_3D_asym})$^{-1}$ & (\ref{transform_1D_asym})$^{-1\ast}$ \\ \cline{2-4}
			& Absorbing CIR
			& (\ref{transform_3D_asym}) & (\ref{transform_1D_asym})$^\ast$ \\ \cline{2-4}
			& Absorption Rate
			& (\ref{transform_3D_instant_asym}) & (\ref{transform_1D_instant_asym}) \\ \hline
		\end{tabular}
	}
	\label{table_equations}
\end{table}

\section{Implementation Details}
\label{sec_implementation}

In this section, we briefly describe how to measure the energy and absorption rate detector signals so that we can verify the transforms for active and passive receivers via simulations. To do so, we need to accommodate discrete (and noisy) observations of the CIRs and not just the analytical expressions in Section~\ref{sec_model}. We assume that, for \emph{either} receiver model, the RX takes $\M$ samples between time $\tx{}=0$ and time $t=\T$. The samples are equally spaced in time by sampling interval $\Delta\tx{\M} = \T/\M$. The individual sample taken at time $\tx{\smM}$ is labeled as $\Nxt{\tx{\smM}}{RX}$, where $\tx{\smM} = \smM\Delta\tx{\M}$, $\smM \in \{1,2,\ldots,\M\}$.

First, we consider the energy detector. Energy detection by a passive RX has been analyzed in \cite{Llatser2013,Meng2014} by integrating the continuous CIR (e.g., (\ref{passive_response_integral})). To accommodate discrete observations, we consider the weighted sum detector that we introduced in \cite[Eq.~(37)]{Noel2014d}, where the weighted sum is
\begin{equation}
\sum_{\smM=1}^{\M}\wx{\smM}\Nxt{\tx{\smM}}{RX},
\end{equation}
and $\wx{\smM}$ is the $\smM$th weight. In \cite{Noel2014d}, we described the equal weight detector, where $\wx{\smM}=1\, \forall \smM$, as analogous to an energy detector. However, the corresponding sum is not precisely an energy detector. To be an energy detector, each sample must also be scaled by the sampling interval $\Delta\tx{\M}$ (as done in \cite{Mahfuz2015}). If not, then we can artificially detect more energy if we take more samples over the same sampling interval. By including the sampling interval, we design the discrete energy detector for the passive RX $\EDt{\cdot}|^\pass$ as
\begin{equation}
\EDt{\tx{\smM}+\frac{\Delta\tx{\M}}{2}}\Big|^\pass = \Delta\tx{\M}\sum_{l=1}^{\smM}\Nxt{\tx{l}}{RX},
\label{energy_detector}
\end{equation}
which applies to \emph{any} environment, and we add the shift $\frac{\Delta\tx{\M}}{2}$ to the energy detector observation time because the $l$th sample accounts for the energy in the continuous interval $[\tx{l}-\frac{\Delta\tx{\M}}{2},\tx{l}+\frac{\Delta\tx{\M}}{2}]$.

The net number of absorbed molecules is used in \cite{Heren2015} as an absorption rate detector, but the number of molecules arriving over the interval $[\tx{\smM},\tx{\smM+1}]$, i.e.,
\begin{equation}
\Nxt{\tx{\smM+1}}{RX} - \Nxt{\tx{\smM}}{RX},
\end{equation}
is only a legitimate rate if we divide the difference by the sampling interval. Therefore, to be consistent with the absorption rate, we design the absorption rate detector $\Delta\Nxt{\cdot}{RX}|^\absorb$ as
\begin{equation}
\Delta\Nxt{\tx{\smM}+\frac{\Delta\tx{\M}}{2}}{RX}\Big|^\absorb = \frac{\Nxt{\tx{\smM+1}}{RX} - \Nxt{\tx{\smM}}{RX}}{\Delta\tx{\M}},
\label{absorption_rate_detector}
\end{equation}
which also applies to \emph{any} environment, and we shift the observation time because (\ref{absorption_rate_detector}) approximates the slope at the midpoint of the interval $[\tx{\smM},\tx{\smM+1}]$.

\section{Simulation and Numerical Results}
\label{sec_results}

In this section, we verify the transform functions derived in Section~\ref{sec_unifying} via simulations and numerical evaluation. We consider passive and absorbing receivers in two systems (one that is 3D and one that is 1D) that are summarized in Table~\ref{table_model_param}.

\begin{table}[!tb]
	\centering
	\caption{Simulation system parameters.}
	
	{\renewcommand{\arraystretch}{1.4}
		\begin{tabular}{l|c|c||c|c}
			\hline
			\bfseries Parameter & \bfseries Symbol & \bfseries Units& \bfseries System 1 & \bfseries System 2 \\ \hline \hline
			\# of Dimensions & - & -
			& 3 & 1 \\ \hline
			RX Radius & $\rrx$ &	$\mu\metre$
			& 1 & 0.5 \\ \hline
			Molecules Released & $\Nx{}$ & $\mol$
			& $10^4$ & $10^3$ \\ \hline
			Distance to RX & $\dist{}$ & $\mu\metre$
			& 5 & 5 \\ \hline
			Diffusion Coeff.
			& $\Dx{}$ & $\metre^2/\second$
			& $10^{-9}$ & $10^{-9}$ \\ \hline
			Sampling Period & $\Delta\tx{\M}$ & $\mu\second$
			& 40 & $2\times10^3$ \\ \hline
			Passive Time Step & $\dt{sim}$ & $\mu\second$
			& 20 & $10^3$ \\ \hline
			Absorbing Time Step & $\dt{sim}$ & $\mu\second$
			& 2 & 10 \\ \hline
			\# of Realizations & - & -
			& $10^4$ & $10^3$ \\ \hline
		\end{tabular}
	}
	\label{table_model_param}
\end{table}

The simulations were completed using release v0.5 of the AcCoRD simulator \cite{Noel2016} and averaged over the number of independent realizations listed in Table~\ref{table_model_param}. The realizations were sufficient to make confidence intervals negligible compared to deviations from the analytical expressions. The simulated 3D environment is truly unbounded, whereas the 1D environment is a $1\,\metre\metre\times1\,\mu\metre\times1\,\mu\metre$ rectangular pipe that is centered at the RX. The simulations of the absorbing RX remove and count molecules if their trajectory in one simulation time step crosses the RX boundary. The time step $\dt{sim}$ for the absorbing RX is small enough to accurately model the absorption. The accuracy of the passive RX is independent of its time step. The sampling period $\Delta\tx{\M}$ is double the time step used in passive RX simulations, so that there are samples at the times shifted by $\frac{\Delta\tx{\M}}{2}$ for the implementation of the energy detector and absorption rate detector in (\ref{energy_detector}) and (\ref{absorption_rate_detector}), respectively.

\begin{figure}[!tb]
	\centering
	\includegraphics[width=\linewidth]{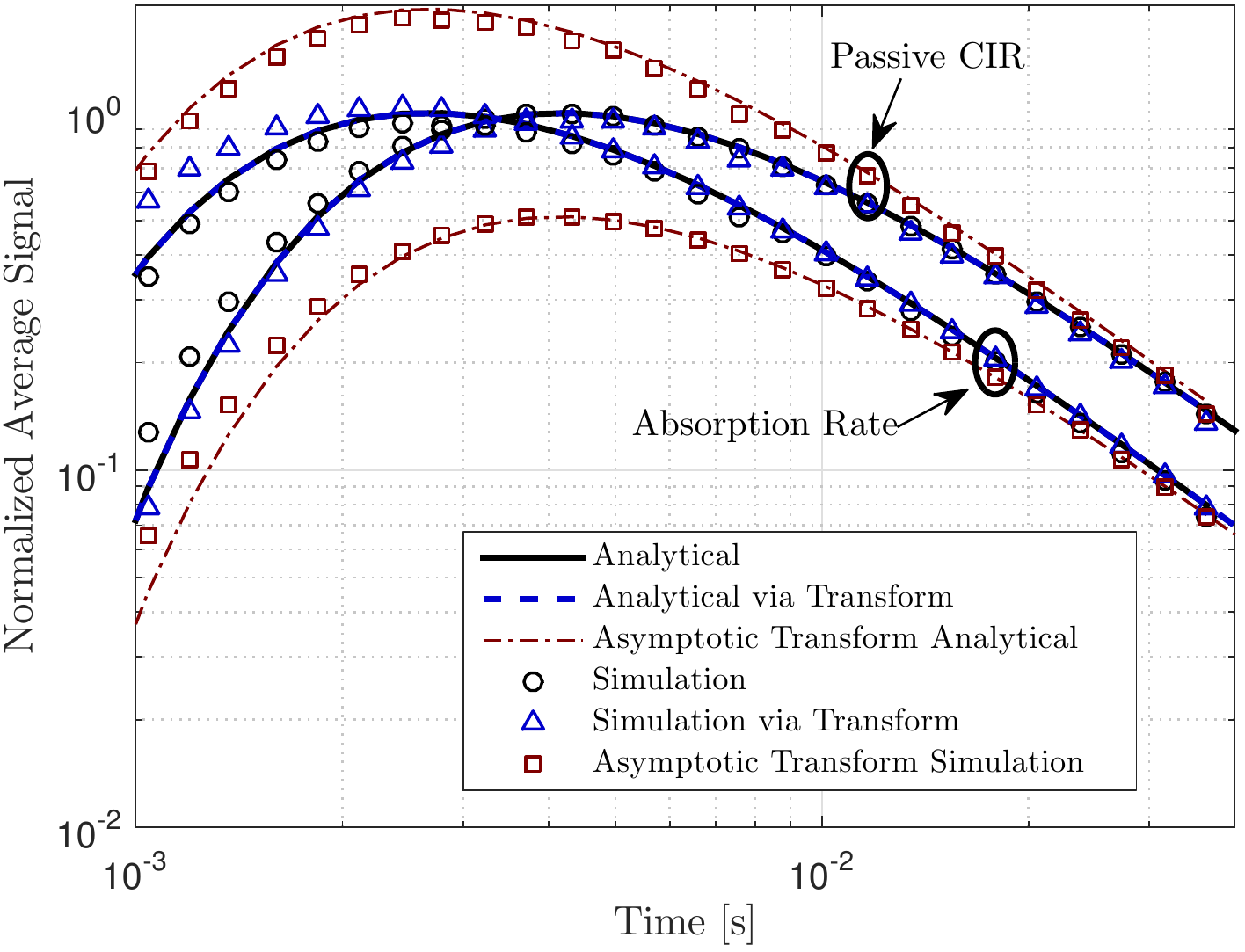}
	\caption{Passive CIR and absorption rate of System 1 (3D) as a function of time. All curves are normalized to the maximum value of the corresponding analytical curve (solid black line), as listed in Table~\ref{table_norm}. The transformed signals are found by applying (\ref{transform_3D_instant}) and (\ref{transform_3D_instant_asym}).}
	\label{fig_3D_CIR}
\end{figure}

\begin{figure}[!tb]
	\centering
	\includegraphics[width=\linewidth]{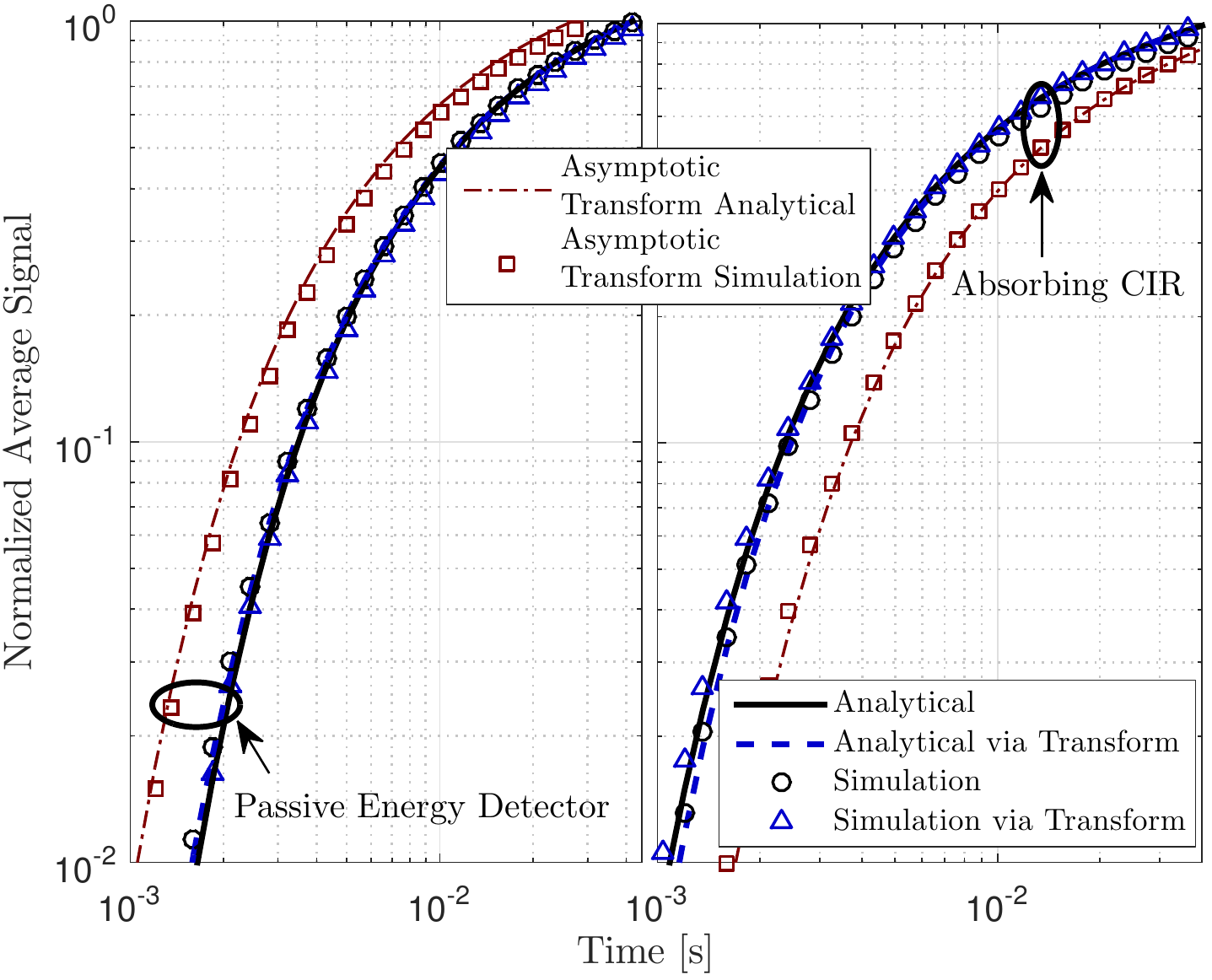}
	\caption{Passive energy detector (left) and absorbing CIR (right) of System 1 (3D). All curves are normalized to the maximum value of the corresponding analytical curve (solid black line), as listed in Table~\ref{table_norm}. The transformed signals are found by applying (\ref{transform_3D}) and (\ref{transform_3D_asym}).}
	\label{fig_3D_ED_Delta}
\end{figure}

\begin{figure}[!tb]
	\centering
	\includegraphics[width=\linewidth]{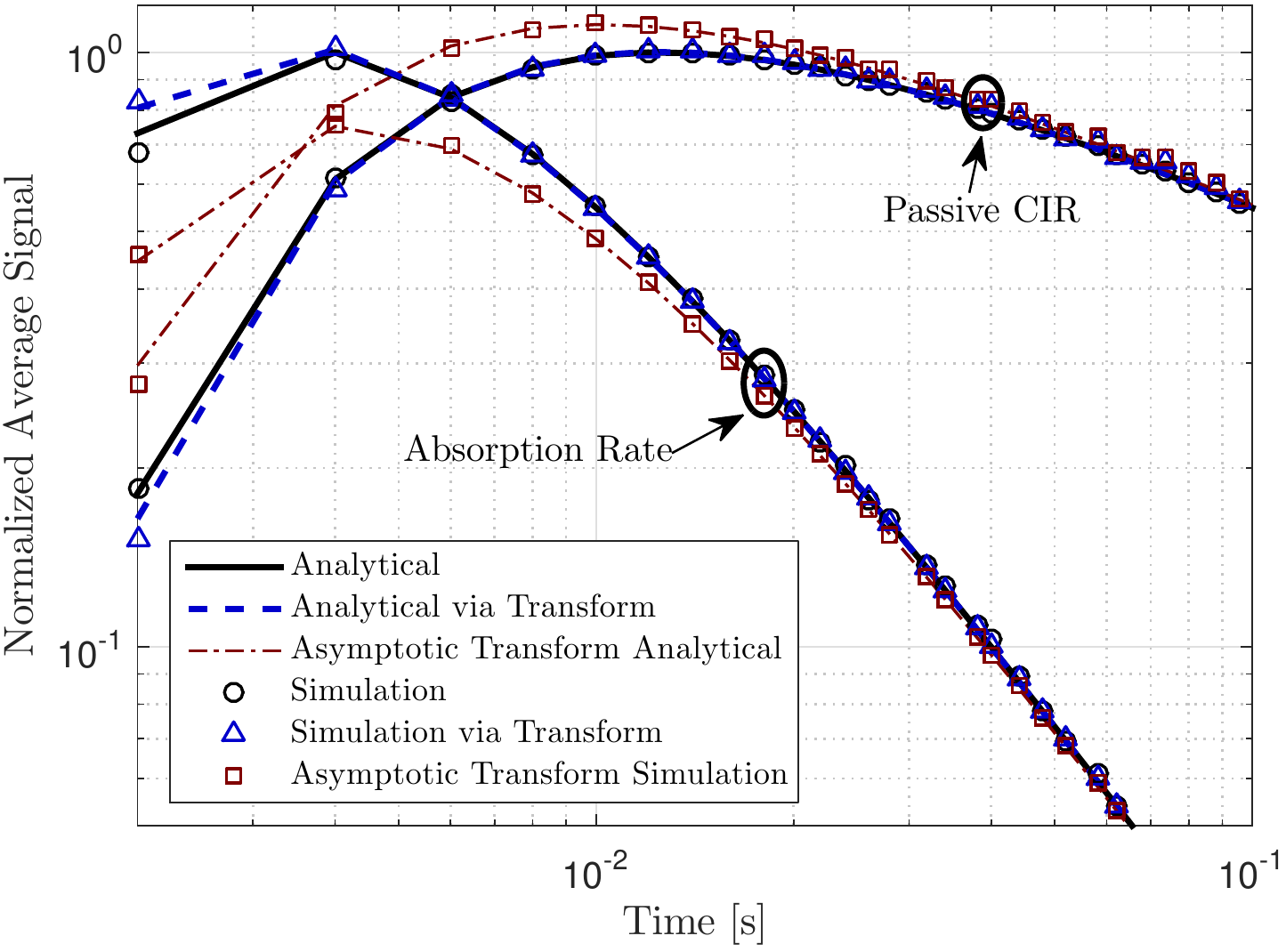}
	\caption{Passive CIR and absorption rate of System 2 (1D) as a function of time. All curves are normalized to the maximum value of the corresponding analytical curve (solid black line), as listed in Table~\ref{table_norm}. The transformed signals are found by applying (\ref{transform_1D_instant}) and (\ref{transform_1D_instant_asym}).}
	\label{fig_1D_CIR}
\end{figure}

The results are presented in Figs.~\ref{fig_3D_CIR}, \ref{fig_3D_ED_Delta}, and \ref{fig_1D_CIR}. They all have the same format and are plotted on a log-log scale so that inaccuracies can be clearly observed. Figs.~\ref{fig_3D_CIR} and \ref{fig_1D_CIR} display the passive CIR and the absorption rate (for Systems 1 and 2, respectively). Fig.~\ref{fig_3D_ED_Delta} displays the passive energy detector and the absorbing CIR for System 1. All curves are grouped according to the signal that they correspond to, and are normalized by the maximum of the corresponding analytical curve (the values of which are listed in Table~\ref{table_norm}). For the specific equation used for each analytical curve, please refer to Table~\ref{table_equations}. The ``Simulation'' curves are analogous to the ``Analytical'' curves, except that the receiver signals (including those used to calculate the energy and the absorption rate detector signals) are replaced by the corresponding average simulation output. Time scales are sufficient to show the asymptotic curves tending toward the time-varying curves.

\begin{table}[!tb]
	\centering
	\caption{Maximum values used to normalize all corresponding curves.}
	
	{\renewcommand{\arraystretch}{1.3}
		\begin{tabular}{l|c|c||c|c}
			\hline
			\bfseries Signal & \bfseries Symbol & \bfseries Units& \bfseries System 1 & \bfseries System 2 \\ \hline \hline
			Passive CIR & $\NxAvg{RX}|^\pass$ & $\mol$
			& 24.7 & 48.4 \\ \hline
			Absorbing CIR & $\NxAvg{RX}|^\absorb$ & $\mol$
			& 1325 & 775 \\ \hline
			Passive ED & $\EDAvg|^\pass$ & $\mol\cdot\second$
			& $1.16\times10^5$ & -\\ \hline
			Absorption Rate & $\Delta\NxAvg{RX}|^\absorb$ & $\frac{\mol}{\second}$
			& 0.390 & - \\ \hline
		\end{tabular}
	}
	\label{table_norm}
\end{table}

All three figures show excellent agreement between the expected signals and those obtained via transforms, both analytically and using the simulations, so we will focus our discussion on a few key observations. First, we comment on the accuracy of the analytical transforms. All non-asymptotic transforms are \emph{very} accurate for the entire range of time. They demonstrate that we have accurate means for transforming between the active and passive spherical receiver models. The accuracy of the asymptotic transforms greatly improves with time. They are not suitable for capturing the signal peak values, but may be appropriate for modeling ISI.

Second, we comment on the simulation accuracy. The accuracy of the absorbing RX signal depends on the simulation time step used. Even though we used time steps that were much smaller than those for the passive RX, the simulations underestimate the absorbing RX signal, particularly for small values of $t$. This can be observed in the simulation of the absorbing signal in all three figures. The simulated passive RX signals and the transforms that depend on those signals show deviations for very small values of $t$ because we assumed that the RX concentration is uniform. However, these simulations become more accurate with increasing $t$. In fact, the transforms of the passive signal become \emph{more} accurate than the absorbing simulation. For example, by $t=0.003\,\second$ in Figs.~\ref{fig_3D_CIR} and \ref{fig_3D_ED_Delta}, the 3D absorbing RX signal is more accurately obtained by transforming the passive simulation than by simulating the absorbing RX directly. If the same time step were used for both receiver models, then this improvement would be much more pronounced. Thus, for this system we can gain in \emph{both} accuracy \emph{and} computational efficiency by simulating a passive RX and transforming the results to apply to an active signal.

\section{Conclusions}
\label{sec_concl}

In this paper, we studied the use of transforms to convert between the observed signal at a passive receiver and that at an absorbing receiver. We used two approaches to derive the transform functions for a spherical receiver in an unbounded diffusion-only system. We modified the weighted sum detector and the hitting rate detector to properly implement an integrator and a differentiator of the passive and absorbing impulse responses, respectively. Numerical results and simulations were shown to verify the transforms. Our future work will consider the derivation of transforms for more complex diffusion-based environments. We will also relax the use of the uniform concentration assumption to improve the accuracy in transforming the passive signal, and will consider the impact of the transforms on the channel statistics.

\bibliography{2016_globe_passive_active}

\end{document}